\documentclass[preprint,12pt]{elsarticle}

\usepackage{amssymb}
\usepackage{mathtools,etoolbox}
\usepackage{caption}
\usepackage{subcaption}
\usepackage{lineno}
\DeclarePairedDelimiterX{\abs}[1]{\lvert}{\rvert}{\ifblank{#1}{{}\cdot{}}{#1}}

\journal{Nuclear Instruments and Methods in Physics Research A}

\begin{document}

\begin{frontmatter}



\title{RPC based tracking system at CERN GIF++ facility}


\author[c]{K. Mota Amarilo}
\ead{kevin.amarilo@cern.ch}
\author[a]{A. Samalan}
\author[a]{M. Tytgat}
\author[aa]{M. El Sawy}
\author[b]{G.A. Alves}
\author[b]{F. Marujo}
\author[b]{E.A. Coelho}
\author[c]{E.M. Da Costa}
\author[c]{H. Nogima}
\author[c]{A. Santoro}
% ...

\affiliation[a]{Ghent University, Dept. of Physics and Astronomy, Proeftuinstraat 86, B-9000 Ghent, Belgium.}
\affiliation[aa]{Université Libre de Bruxelles, Avenue Franklin Roosevelt 50-1050 Bruxelles, Belgium.}
\affiliation[b]{Centro Brasileiro Pesquisas Fisicas, R. Dr. Xavier Sigaud, 150 - Urca, Rio de Janeiro - RJ, 22290-180, Brazil.}
\affiliation[c]{Dep. de Fisica Nuclear e Altas Energias, Instituto de Fisica, Universidade do Estado do Rio de Janeiro, Rua Sao Francisco Xavier, 524, BR - Rio de Janeiro 20559-900, RJ, Brazil.}
% ...

\begin{abstract}
With the HL-LHC upgrade of the LHC machine, an increase of the instantaneous 
luminosity by a factor of five is expected and the current detection systems need to be validated for such working conditions to ensure stable data taking. At the CERN Gamma Irradiation Facility (GIF++) many muon detectors undergo such studies, but the high gamma background can pose a challenge to the muon trigger system which is exposed to many fake hits from the gamma background. A tracking system using RPCs is implemented to clean the fake hits, taking profit of the high muon efficiency of these chambers. This work will present the tracking system configuration, used detector analysis algorithm and results.
\end{abstract}



\begin{keyword}
Resistive plate chambers \sep CMS Experiment \sep gaseous detectors \sep HL-LHC 
\PACS 0000 \sep 1111
\MSC 0000 \sep 1111
\end{keyword}

\end{frontmatter}
\section{Introduction}

The Resistive Plate Chambers (RPC) act as muon triggering detector in the Muon System of the Compact Muon Solenoid (CMS) experiment \cite{CMS:2008xjf}. With the coming upgrade of the Large Hadron Collider (LHC) machine, the High Luminosity LHC (HL-LHC) \cite{Aberle:2749422}, many detector systems are preparing to improve their capabilities. In particular, the CMS RPC project has two important planned activities -- the replacement of the off-detector electronics (called link system) and the extension of the RPC coverage from $|\eta| = $ 1.9 to 2.4 with the new improved RPC (iRPC) detectors \cite{CERN-LHCC-2017-012}.

To study the performance and stability of RPCs in the HL-LHC conditions, tests are taking place in the CERN Gamma Irradiation Facility (GIF++), where a high energy particle beam (normally muons) and photons from a gamma source (13 TBq, $^{137}$Cs) are combined \cite{guida2015gif++}. The maximum expected background rate for the RPCs in the current CMS Muon system is 600 Hz/cm$^2$ and for the new iRPCs it goes up to 2 kHz/cm$^2$ (including the safety factor of three) \cite{CERN-LHCC-2017-012}. Such high rates can pose a challenge to the measurement, as the fake hits can bias the data-taking. 

All results shown in this work were recorded during GIF++ test beam in October 2021.

\section{Experimental Setup}

\begin{figure}[htbp!]
  \centering
  \includegraphics[width=.7\textwidth]{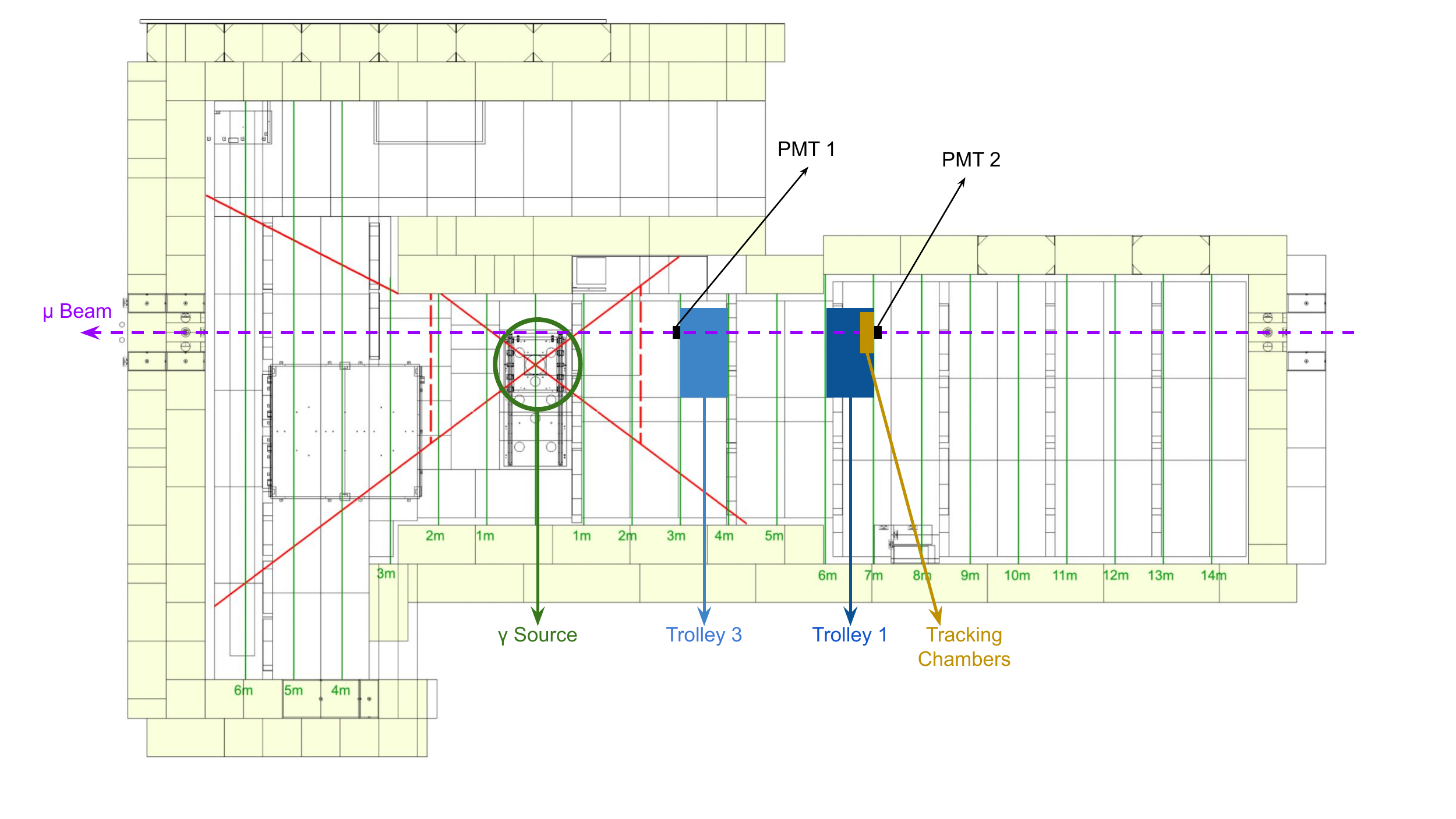}
  \caption{Experimental setup at GIF++.}
  \label{fig:gif_setup}
\end{figure}

Figure \ref{fig:gif_setup} shows a representation of the setup in GIF++. There are two trolleys where the RPC detectors are placed for irradiation. In addition to the RPCs, two scintillators are placed on each trolley to trigger on the muon beam. The data acquisition is performed using a CAEN Time-to-Digital Converter (TDC) module of type V1190A where the frontend electronics of the chambers are connected. A V1718 VME controller module is responsible for the communication between the TDC and the computer where the data are stored. To host the VME controller and the TDC a 6U VME 6021 crate is used.

In the trolley farther from the gamma source, there are two tracking RPC chambers with their strip planes oriented perpendicular to each other to allow measurement in two directions. The chambers used are the following:
\begin{itemize}
  \item Tracking chambers: Referred to as GT1 and GT2, these are double gap chambers made of 2 mm thick High Pressure Laminate (HPL, also known as bakelite). They are equipped with a strip plane with 32 strips of 1.45 cm pitch. The signals are read with the CMS
  FrontEnd electronics with threshold set to 150 fC.
  \item Test Chamber: Referred to as KODEL-C, this is also a double gap chamber made of   HPL with 1.4 mm thickness. It is equipped with a strip plane with 32 strips of   1.95 cm pitch and custom FrontEnd electronics with threshold set to 75 fC.
\end{itemize}

All the chambers are flushed with the CMS RPC standard gas mixture (95.2\% freon ($C_2H_2F_4$), 4.5\% isobutane ($iC_4H_{10}$), and 0.3\% sulphur hexafluoride ($SF_6$)) with a relative humidity of $\approx$40\%. The tracking chambers are always powered on with the high voltage (HV) working point (WP) applied to their electrodes.

\section{Tracking algorithm}

\begin{figure}[htbp!]
  \centering
  \includegraphics[width=.7\textwidth]{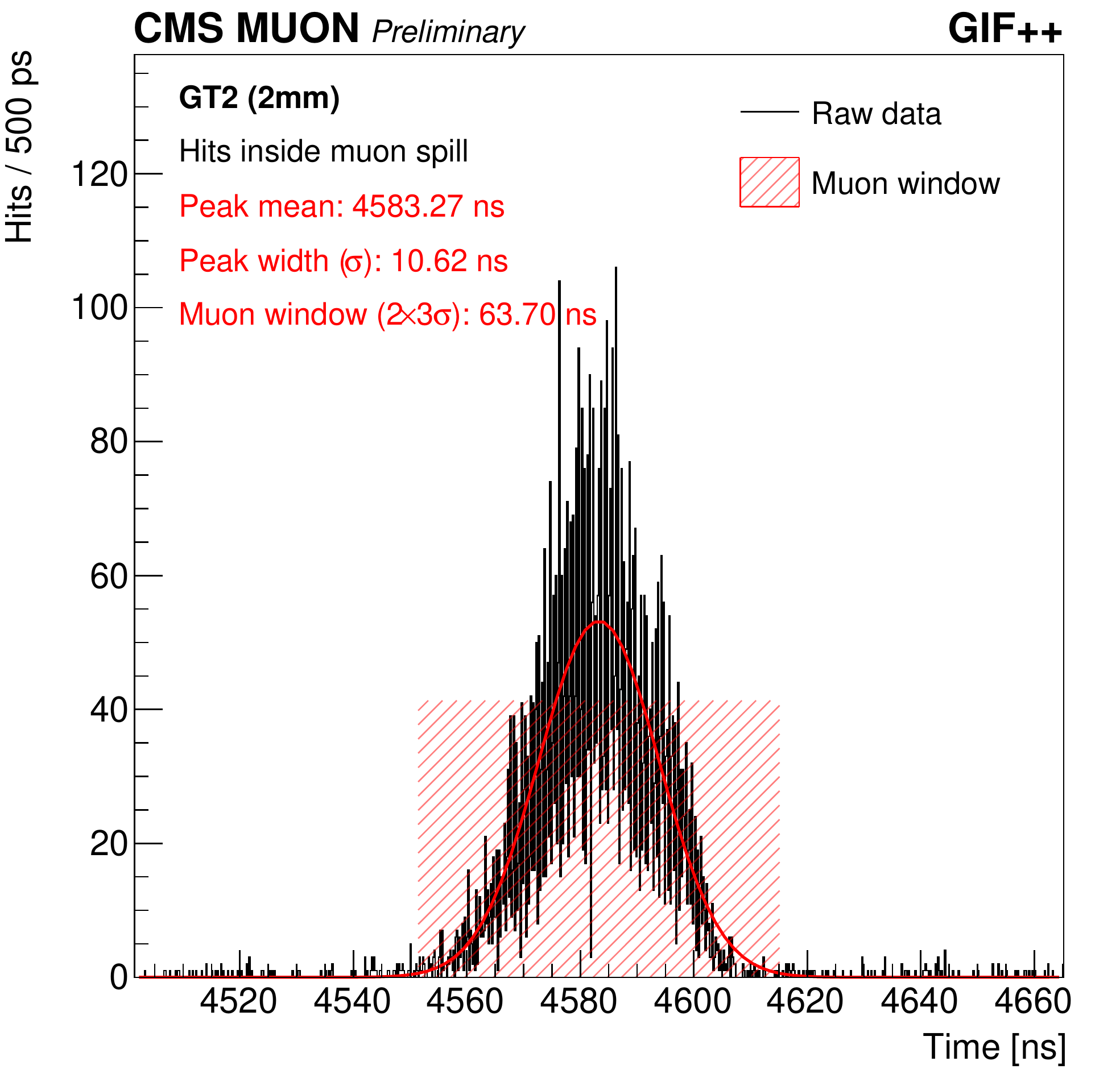}
  \caption{Hits time profile of the tracking chamber GT2. The time window where the hits are accepted in the analysis is represented in the red hatched box.}
  \label{fig:timeprofile}
\end{figure}

\begin{figure}[htbp!]
  \centering
  \includegraphics[width=.7\textwidth]{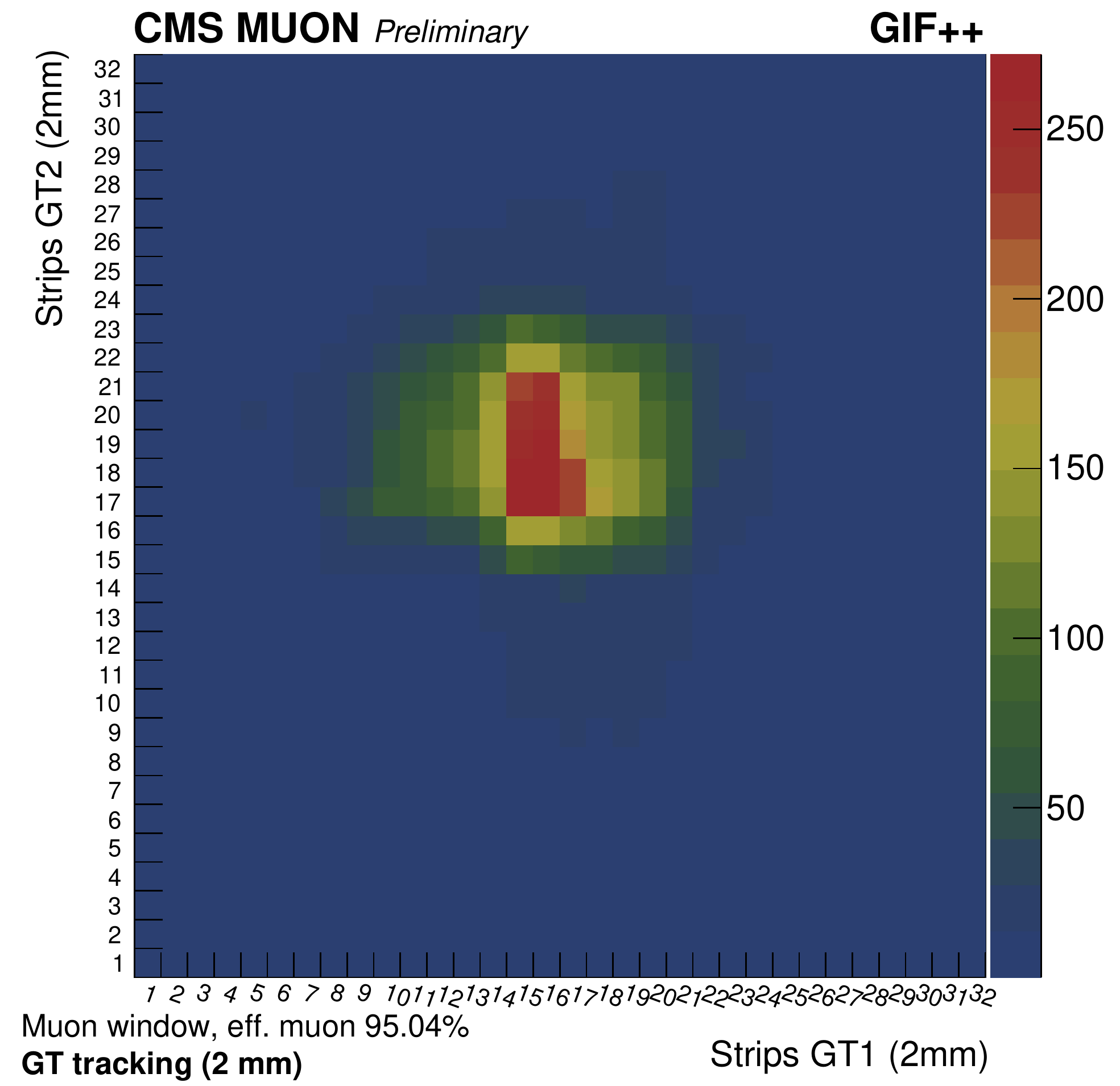}
  \caption{2D hit profile of the strips signals for the hits in the muon time window.}
  \label{fig:2Dtimeprofile}
\end{figure}

To enable the tracking analysis, at least one hit is required in both tracking chambers inside the muon time window defined by a Gaussian fit, as indicated in Fig. \ref{fig:timeprofile}. A 2D hit profile is made to check their alignment with the muon beam, as shown in Fig. \ref{fig:2Dtimeprofile}. The hit profile is also determined for the test chamber and it is used for alignment between the chambers.

The tracking algorithm relies on the assumption that the beam is perpendicular to the strip plane, therefore it is only needed to extrapolate the position of the hit from the tracking to the test chamber and check for a matching hit. For every event, the following steps are taken:
\begin{enumerate}
    \item Perform the clusterization of the hits in the tracking chambers, where 
    events with more than one cluster are rejected. The 
    cluster barycentres are defined as the mean position in the cluster;
    \item Perform the clusterization for the test chambers and calculate the 
    clusters' barycentres;
    \item Form a perpendicular track starting from the tracking cluster barycentre;
    \item Check for a match in any cluster on the test chamber. 
\end{enumerate}

\begin{figure}[htbp!]
\centering 
\includegraphics[width=.7\textwidth]{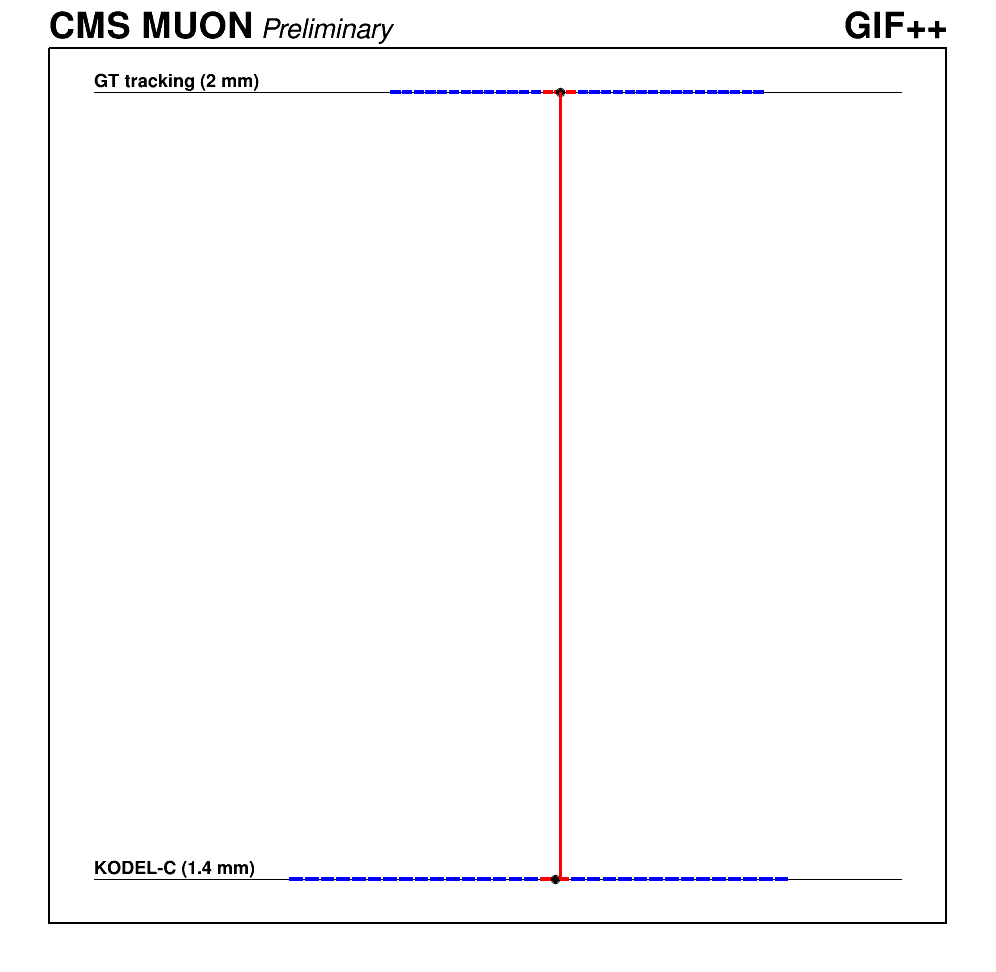}
\caption{Example of one event in which the hit on tracking chamber was matched on the test chamber. The strips in red represent the hits and the point in black is the 
cluster barycentre.}
\label{fig:evt}
\end{figure}

Figure \ref{fig:evt} shows an example of event where the analysis was performed and a matching hit was found.

\section{Performance of the Tracking Algorithm}

\begin{figure}[htbp!]
  \centering
  \begin{subfigure}[b]{0.45\textwidth}
      \centering
      \includegraphics[width=\textwidth]{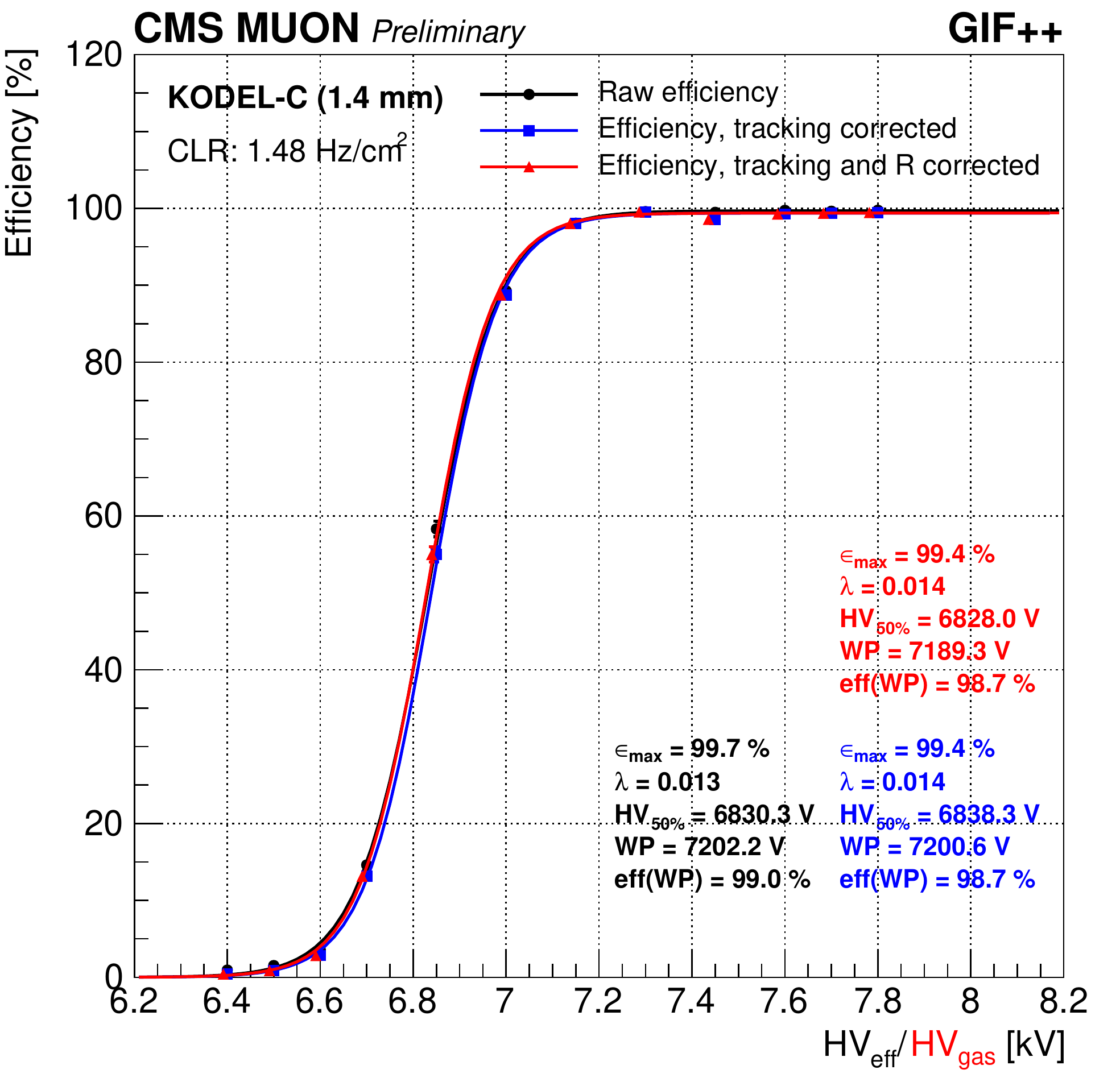}
      \caption{}
      \label{subfig:sourceOFF}
  \end{subfigure}\hfil
  \begin{subfigure}[b]{0.45\textwidth}
      \centering
      \includegraphics[width=\textwidth]{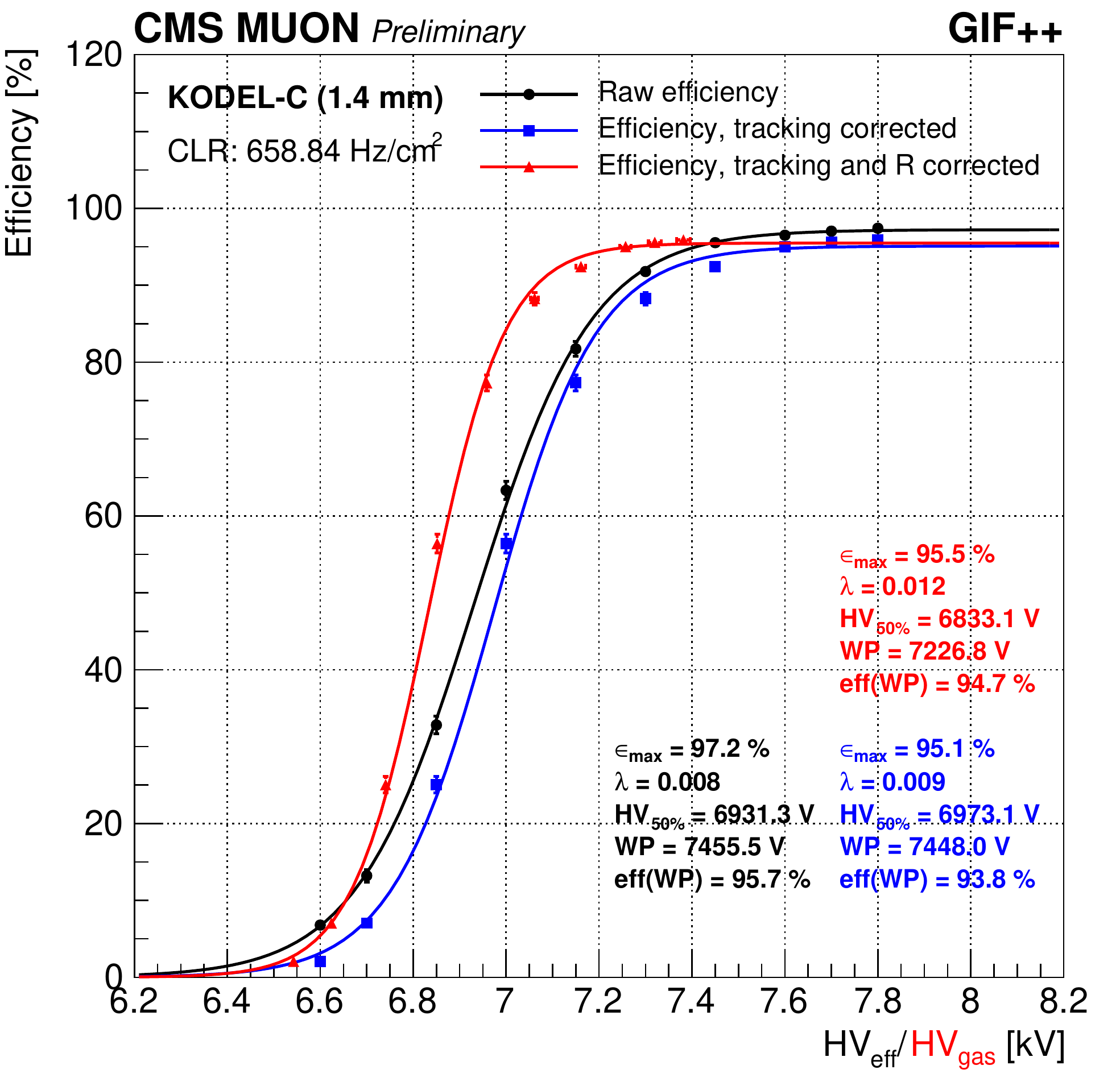}
      \caption{}
      \label{subfig:600Hz}
  \end{subfigure}\hfil
  \begin{subfigure}[b]{0.45\textwidth}
      \centering
      \includegraphics[width=\textwidth]{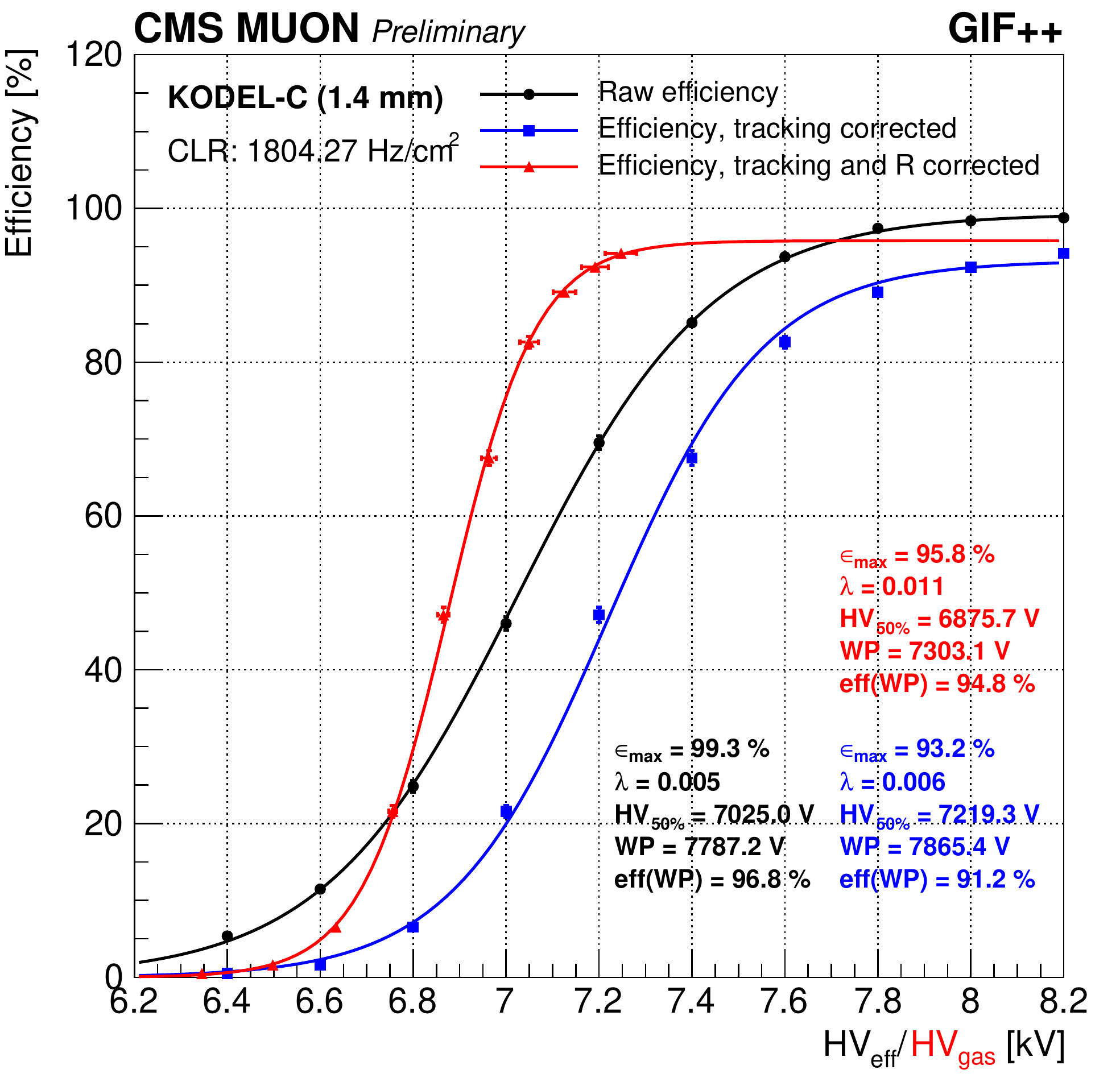}
      \caption{}
      \label{subfig:1800Hz}
  \end{subfigure}
  \caption{Efficiencies and their sigmoid fits measured at three photon fluxes with measured at the working-point high voltages gamma cluster rates of 1.804 kHz (a), 0.645 kHz (b), and 1.48 Hz (c), respectively. The curve in black is the efficiency calculated with all hits inside the muon window, the blue one is with applied tracking correction and the one in red -- with tracking and resistance correction.}
  \label{fig:effcomparison}
\end{figure}

Efficiency curves of the test chambers were used to evaluate the validity of the tracking in rejecting the fake hits caused by the gammas. On Fig. \ref{fig:effcomparison} the efficiency curves are compared for three gamma background conditions. There are three curves in each plot. The black curve is calculated without the tracking, taking into consideration all the hits in the muon time window. For the blue one, only the hits that passed the tracking criteria were considered. Finally, for the red curve, the HV was recalculated to remove the voltage drop caused by the resistance of the electrodes - this correction decouples the shift to the right of the curve on higher cluster rates, caused by the increase of the current. Therefore, HV$_{gas}$ is the effective HV applied to the gas volume. 

As expected, in Fig. \ref{subfig:sourceOFF}, where the data were taken in a run with source OFF, the three curves are equivalent since the cluster rate (CLR) and the currents are low. in Fig. \ref{subfig:600Hz} with CLR $\approx$ 648 Hz/cm$^2$, a shift to the right on the tracking corrected curve is observed, and it is much more evident in Fig. \ref{subfig:1800Hz} with CLR $\approx$ 1.8 kHz/cm$^2$. The increase of the maximum efficiency on the raw efficiency curve from the Fig. \ref{subfig:600Hz} to \ref{subfig:1800Hz} indicates high gamma contamination.

\begin{figure}
    \centering
    \begin{subfigure}{0.45\textwidth}
        \centering
        \includegraphics[width=\textwidth]{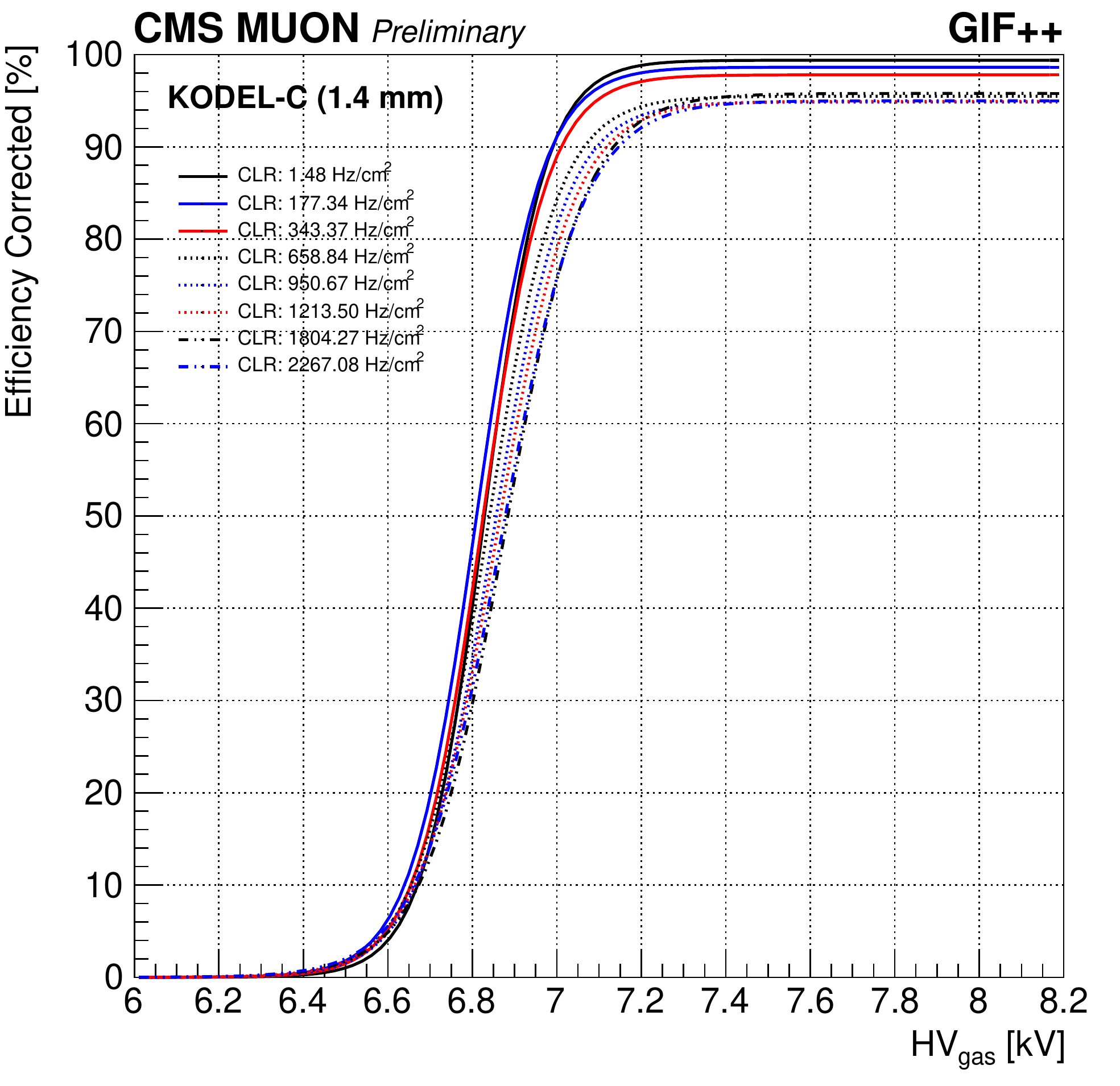}
        \caption{}
        \label{subfig:sigmoids_trk_corr}
    \end{subfigure}\hfil
    \begin{subfigure}{0.45\textwidth}
        \centering
        \includegraphics[width=\textwidth]{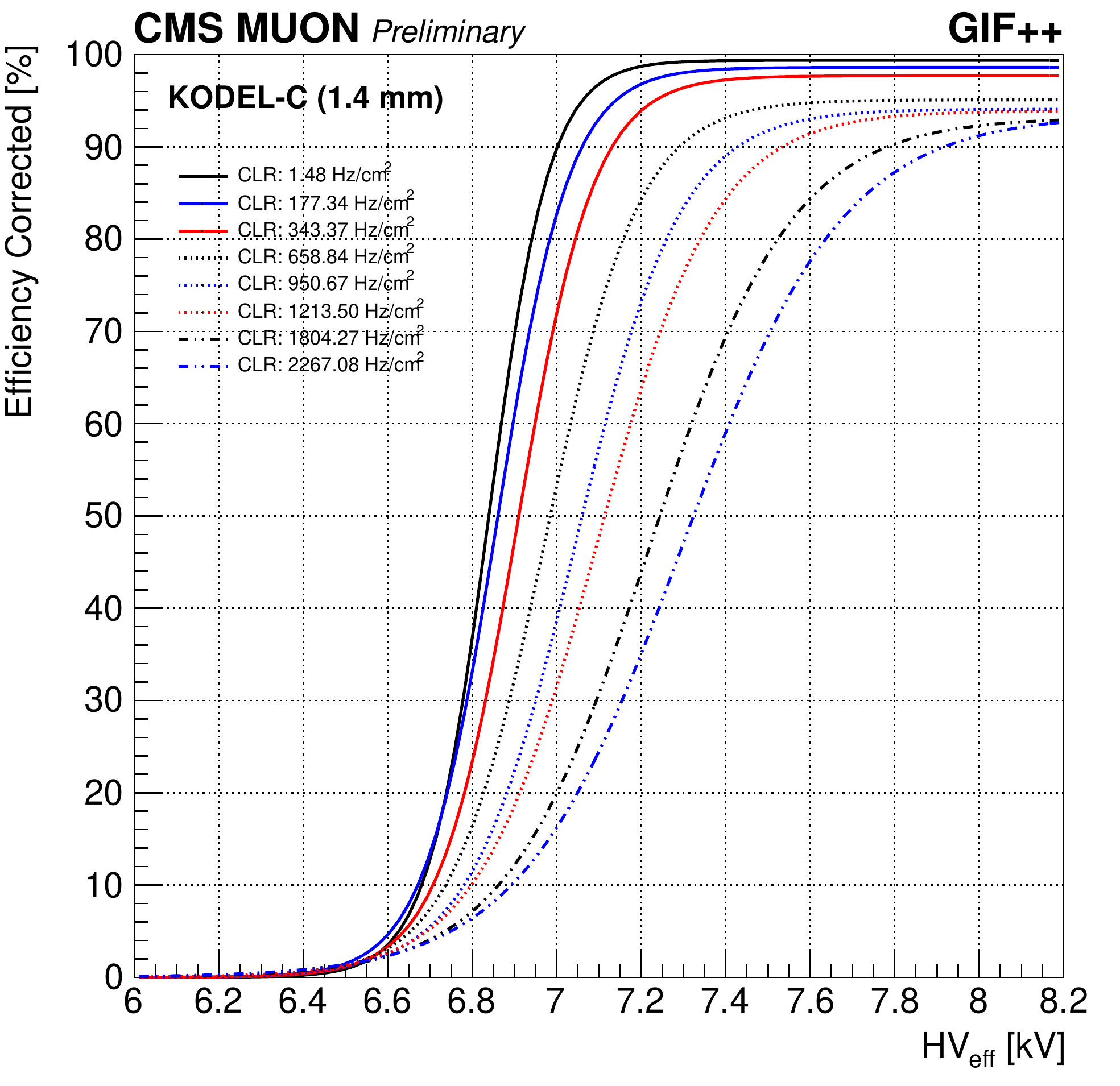}
        \caption{}
        \label{subfig:sigmoids_trk}
    \end{subfigure}
    \caption{Efficiency curves for various gamma background rates. In (a) the HV applied is corrected only by pressure and temperature (HV$_{eff}$), while in (b) the HV is also corrected by the resistance of the gaps (HV$_{gas}$)}.
    \label{fig:effcomparison2}
\end{figure}

In Fig. \ref{subfig:sigmoids_trk_corr}, it is possible to see the curves with tracking and resistance correction. The shift to the right is caused by the reduction of the gas gain at higher rates, while the drop in the maximum efficiency 
is caused by the dead-time of the electronics. In Fig. \ref{subfig:sigmoids_trk}, no resistance correction is applied, so the main reason for the shift is the voltage drop over the electrode's resistance, coupled with the reasons mentioned before.

\section{Conclusion}

The implementation of the tracking system was motivated by the necessity to test the CMS RPC chambers at the conditions of the HL-LHC. The results show that the tracking system performs very well to remove fake hits from the gamma background, even at rates as high as 2 kHz/cm$^2$. The test chamber performed very well and showed increase on the working point of $\approx$650 V with efficiency loss of $\approx$ 7.5\%, using the custom electronics with threshold of 75 fC. This system is currently being used for the ageing studies of the CMS RPC system.

\section*{Acknowledgements}
We would like to thank our colleagues from CERN Gamma Irradiation facility. We would also like to acknowledge the enduring support for the Upgrade of the CMS detector and the supporting computing infrastructure provided by the following funding agencies: FWO (Belgium); CNPq, CAPES and FAPERJ (Brazil); MES and BNSF (Bulgaria); CERN; CAS, MoST, and NSFC (China); MINCIENCIAS (Colombia); CEA and CNRS/IN2P3 (France); SRNSFG (Georgia); DAE and DST (India); IPM (Iran); INFN (Italy); MSIP and NRF (Republic of Korea); BUAP, CINVESTAV, CONACYT, LNS, SEP, and UASLP-FAI (Mexico); PAEC (Pakistan); DOE and NSF (USA).

\appendix

 \bibliographystyle{elsarticle-num} 
 \bibliography{cas-refs}

\begin{thebibliography}{1}
\expandafter\ifx\csname url\endcsname\relax
  \def\url#1{\texttt{#1}}\fi
\expandafter\ifx\csname urlprefix\endcsname\relax\def\urlprefix{URL }\fi
\expandafter\ifx\csname href\endcsname\relax
  \def\href#1#2{#2} \def\path#1{#1}\fi

\bibitem{CMS:2008xjf}
{CMS Collaboration}, {The CMS Experiment at the CERN LHC}, JINST 3 (2008)
  S08004.
\newblock \href {https://doi.org/10.1088/1748-0221/3/08/S08004}
  {\path{doi:10.1088/1748-0221/3/08/S08004}}.

\bibitem{Aberle:2749422}
O.~Aberle, et~al., \href{https://cds.cern.ch/record/2749422}{{High-Luminosity
  Large Hadron Collider (HL-LHC): Technical design report}}, CERN Yellow
  Reports: Monographs, CERN, Geneva, 2020.
\newblock \href {https://doi.org/10.23731/CYRM-2020-0010}
  {\path{doi:10.23731/CYRM-2020-0010}}.
\newline\urlprefix\url{https://cds.cern.ch/record/2749422}

\bibitem{CERN-LHCC-2017-012}
{CMS Collaboration}, \href{https://cds.cern.ch/record/2283189}{{The Phase-2
  Upgrade of the CMS Muon Detectors}}, Tech. rep., CERN, Geneva (2017).
\newline\urlprefix\url{https://cds.cern.ch/record/2283189}

\bibitem{guida2015gif++}
R.~Guida, {GIF++}: The new {CERN} irradiation facility to test large-area
  detectors for {HL-LHC}, in: 2015 IEEE Nuclear Science Symposium and Medical
  Imaging Conference (NSS/MIC), IEEE, 2015, pp. 1--4.

\end{thebibliography}





\end{document}